\documentclass[twocolumn,prb, showpacs]{revtex4}
\usepackage{graphics,graphicx}
\begin{document}
\title{Mott insulator-to-metal transition in yttrium-doped CaIrO$_{3}$}
\author{J.~Gunasekera$^{1}$}
\author{Y.~Chen$^{1}$}
\author{J. W.~Kremenak$^{1}$}
\author{P. F.~Miceli$^{1}$}
\author{D. K.~Singh$^{1,*}$}
\affiliation{$^{1}$Department of Physics and Astronomy, University of Missouri, Columbia, MO 65211}
\affiliation{$^{*}$Email: singhdk@missouri.edu}

\begin{abstract}

We report on the study of insulator-to-metal transition in post-perovskite compound CaIrO$_{3}$. It is discovered that a gradual chemical substitution of calcium by yttrium leads to the onset of strong metallic behavior in this compound. This observation is in stark contrast to BaIrO$_{3}$, which preserves its Mott insulating behavior despite excess of charge carriers due to yttrium doping. Magnetic measurements reveal that both compounds tend to exhibit magnetic character irrespective of the chemical substitution of Ca or Ba. We analyze these unusual observations in light of recent researches that suggest that CaIrO$_{3}$ does not necessarily possess $j$ = 1/2 ground state due to structural distortion. The insulator-to-metal transition in CaIrO$_{3}$ will spur new researches to explore more exotic ground state, including superconductivity, in post-perovskite Mott insulators.

\end{abstract}

\pacs{75.70.Tj, 72.90.+y, 75.47.Lx} \maketitle

Iridium oxide based compounds are generating strong research interest because of intriguing physical properties, arising due to incorporation of the spin-orbit coupling to the Heisenberg Hamiltonian, that includes unusual Mott insulating state and (possible) Kitaev spin-liquid state.\cite{Pesin,Kim,Chaloupka, Laguna,Jackeli} One recent proposal has suggested the application of 5$d$ iridium oxide compound in the spin-current detection.\cite{Fujiwara} The ground state in majority of these materials, $A$$_{2}$IrO$_{3}$ and $B$IrO$_{3}$ (where $A$ = alkali metals and $B$ = alkaline earth metals), is often described as a Mott insulting state, where the Ir$^{4+}$ ion with a ($t$$_{2g}$)$^{5}$ electronic configuration splits into a fully occupied $j$ = 3/2 state and a half-filled $j$ = 1/2 states.\cite{Ashvin,Yogesh,Cao2,Ohgushi1,Maiti} While the quantum-mechanical nature of the spin-orbit coupling is considered key to the quantum Hall effect in topological insulating materials,\cite{Ken} its interaction with cubic crystal field in Ir 5$d$ transition element ensures a $j$ = 1/2 ground state in Mott insulator iridates.\cite{Ohgushi2,Ohgushi1,Sala2} The bandwidth of the $j$ = 1/2 state $w$~$\simeq$~$Nt$, where $N$ is the coordination number and $t$ is the Ir-Ir hopping matrix, is found to be comparable to the Coulomb repulsion potential U in $B$IrO$_{3}$; thus on the verge of attaining the metallic character.\cite{Bogdanov} A simple modification in the structural properties or, the change in the hopping integral achieved via higher carrier density can, therefore, induce a highly desirable metallic behavior in iridium oxide compounds of 1-1-3 phase.

In this letter, we present new results on the presence (absence) of insulator-to-metal transition in the hole-doped CaIrO$_{3}$ (BaIrO$_{3}$). In a novel observation, it is found that the yttrium substitution of calcium leads to an onset of metallic behavior at x = 0.4 in Y$_{x}$Ca$_{1-x}$IrO$_{3}$, which is not the case in Y$_{x}$Ba$_{1-x}$IrO$_{3}$. Both compounds, however, preserve magnetic characteristic irrespective of the chemical doping. In another notable observation, the dimensionality analysis of the electrical transport data below the magnetic transition demonstrates a quasi-2D electronic pattern (d = 2.4) in CaIrO$_{3}$, which is in stark contrast to 3D electronic transport in BaIrO$_{3}$. CaIrO$_{3}$ and BaIrO$_{3}$ manifest Mott insulating behavior with charge gap of $\simeq$~0.17 eV and $\simeq$~50 meV, respectively.\cite{Sala1,Maiti} While CaIrO$_{3}$ crystallizes primarily in the orthorombic phase of crystallographic group $Cmcm$ (similar to MgSiO$_{3}$),\cite{Murakami} BaIrO$_{3}$ forms a monoclinic structure ($C2/m$).\cite{Cao2} In both compounds, the chemical structure is composed of edge-sharing IrO$_{6}$ octahedra along $a$- and $c$- directions, respectively. In CaIrO$_{3}$, however, each octahedron is compressed along the oxygen corner-sharing $c$-direction, which is different from BaIrO$_{3}$ where the monoclinic distortion generates twisting and buckling of the Ir$_{3}$O$_{12}$ trimers that form marginally tilted one-dimensional columns parallel to the $c$-axis. Apparently, the chemical structure plays important roles in their magnetic response-- while CaIrO$_{3}$ is an antiferromagnet, $T$$_{N}$$\simeq$110 K,\cite{Ohgushi1,Ohgushi2} with magnetic moments (canted along $b$-direction) are arranged ferromagnetically along $a$-direction and antiferromagnetically along $c$-direction, BaIrO$_{3}$ exhibits weak ferromagnetism (coexisting with charge density wave) below $T$$_{c}$$\simeq$175 K.\cite{Lindsay,Cao2} It is counter-intuitive to find that the electrical properties of ferromagnetic BaIrO$_{3}$ remains unaffected to the presence of additional charge carriers (hole) while antiferromagnetic CaIrO$_{3}$ exhibits the insulator-to-metal transition as a function of chemical doping percentage. In a most recent report on the resonant inelastic X-ray scattering on CaIrO$_{3}$, it is suggested that the ground state in this compound is not a $j$$_{eff}$ = 1/2 state.\cite{Sala1} All of these unusual phenomana are, perhaps, related to one another with strong link to the underlying chemical structure properties. 

\begin{figure}
\centering
\includegraphics[width=8.5 cm]{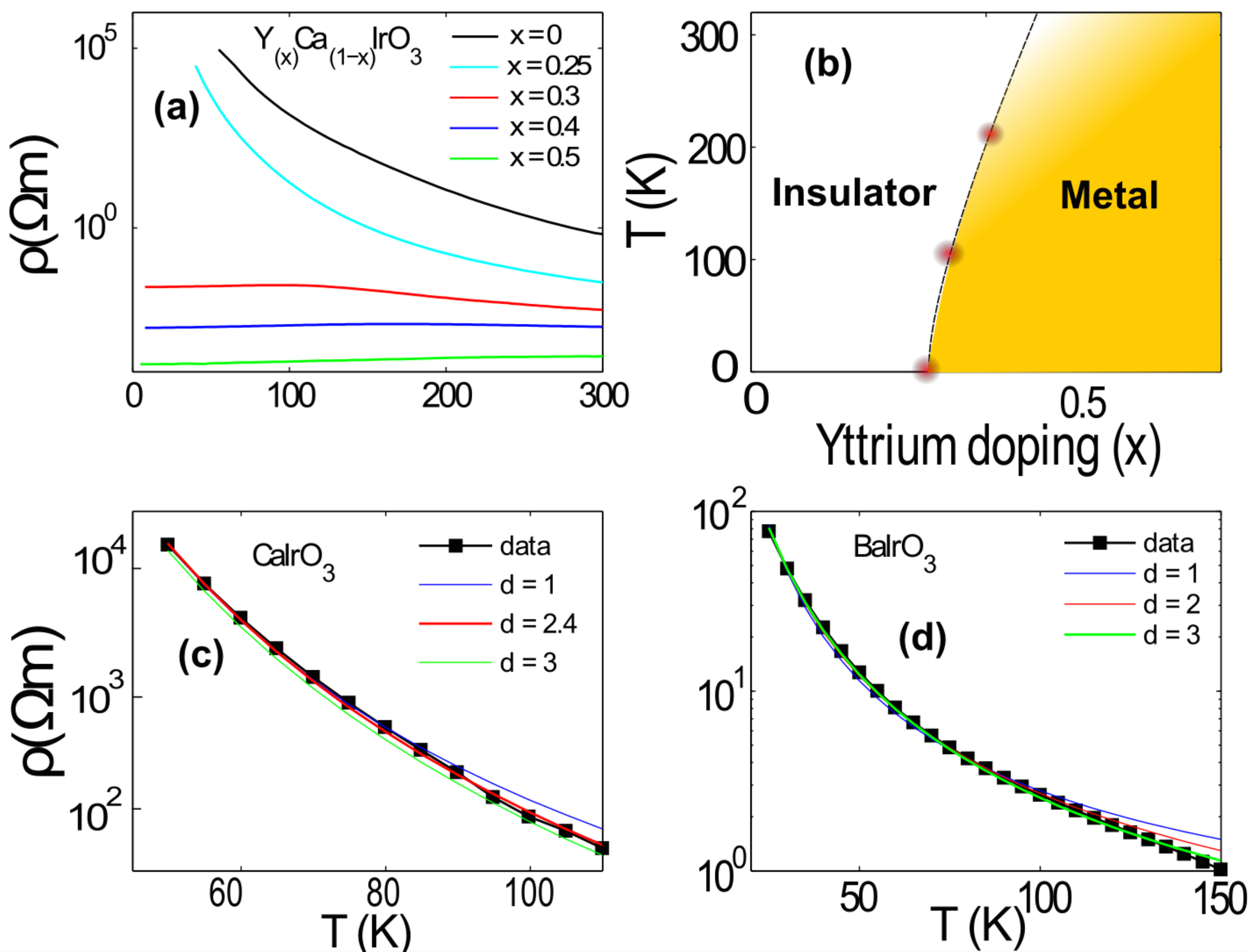} \vspace{-4mm}
\caption{(color online) Phase-diagram of insulator-to-metal transition in yttrium doped CaIrO$_{3}$ (a) Electrical resistivity as a function of temperature for various chemical doping percentage of Y$_{x}$Ca$_{1-x}$IrO$_{3}$. Clearly, a plateau is observed below $T$ $\simeq$ 120 K for $x$ = 0.3. Further substitution of Ca by Y leads to an onset of insulator-metal transition in this system. (b) Qualitative phase-diagram indicating insulator-metal transition as a function of yttrium chemical doping percentage in Y$_{x}$Ca$_{1-x}$IrO$_{3}$. (c) Fitting of electrical resistivity data, below $T$$_{o}$ $\simeq$ 110 K, using variable range hopping model (see text for detail). Best fit of the data is obtained for $d$ = 2.4, thus indicating fractional electronic dimension in this system. For comparison purposes, plots for integer dimensions are also illustrated. (d) Similar analysis for BaIrO$_{3}$ reveals three dimensional electronic pattern.
} \vspace{-4mm}
\end{figure}

The high purity polycrystalline samples of Y$_{x}$Ca$_{1-x}$IrO$_{3}$ and Y$_{x}$Ba$_{1-x}$IrO$_{3}$ were synthesized by conventional solid state reaction method using ultra-pure ingredients of IrO$_{2}$, CaCO$_{3}$, BaCO$_{3}$ and Y$_{2}$O$_{3}$. Starting materials were mixed in stoichiometric compositon, with ten percent extra CaCO$_{3}$ (or, BaCO$_{3}$) to compensate for their rapid evaporation, palletized and sintered at 950$^{o}$ for three days. The furnace cooled samples were grinded, palletized and sintered at 1000$^{o}$ for another three days. Resulting samples were characterized using Siemens D500 powder X-ray diffractometer and Rigaku high resolution X-ray diffractometer, confirming the single phase of materials. Detailed electrical and magnetic measurements on pallets of polycrystalline samples were performed using Quantum Design Physical Properties Measurement System with a temerature range of 2 K - 350 K and magnetic field application range of 7 T. Four probe technique was employed to elucidate electrical properties of Y$_{x}$Ca$_{1-x}$IrO$_{3}$ and Y$_{x}$Ba$_{1-x}$IrO$_{3}$.

\begin{figure}
\centering
\includegraphics[width=9 cm]{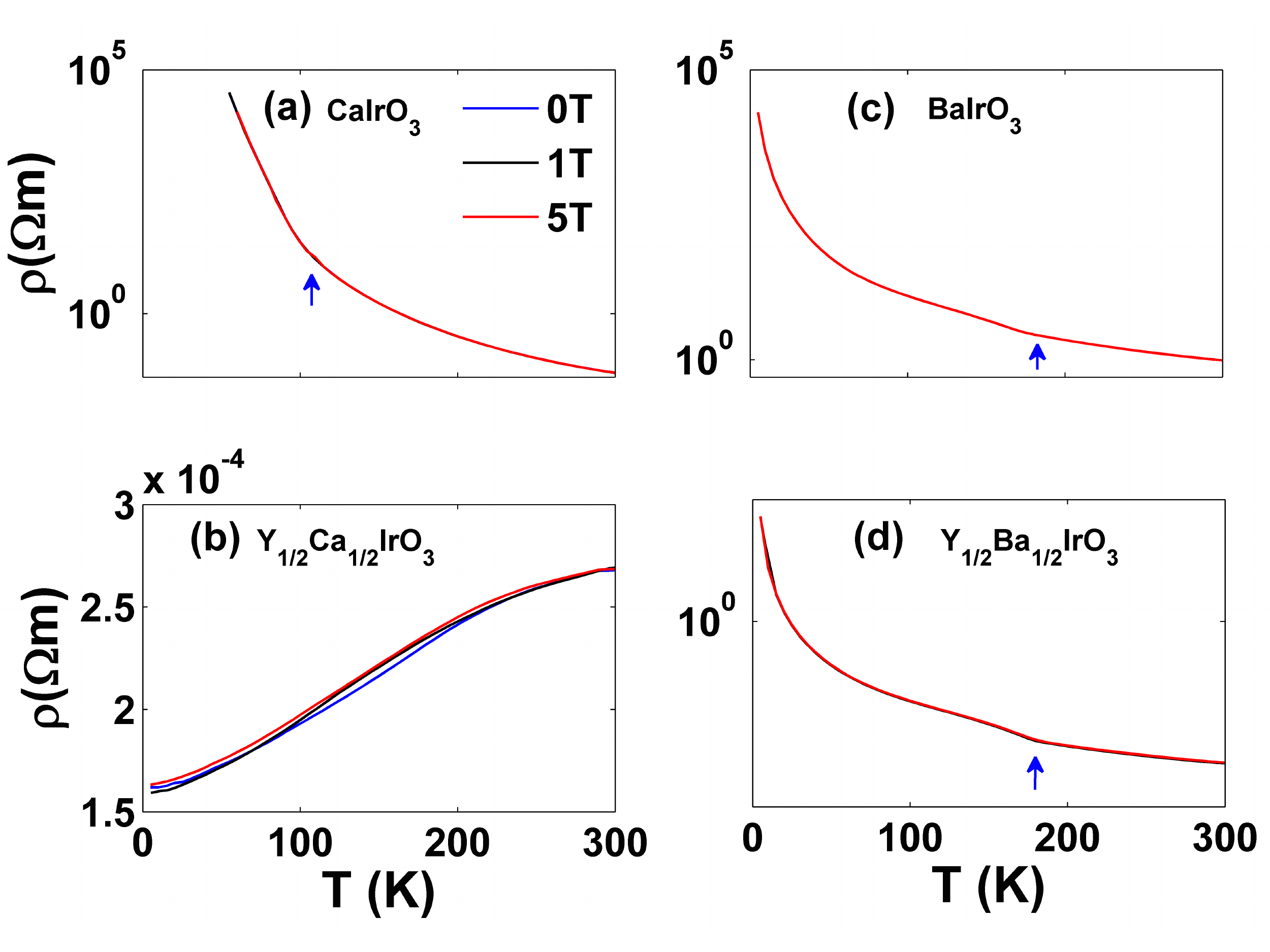} \vspace{-4mm}
\caption{(color online) Electrical transport data in applied magnetic field for two characteristic chemical doping percentages, x = 0 and 0.5. (a) Characteristic plots of electrical resistivity as a function of temperature of CaIrO$_{3}$ at few different values of applied magnetic field are shown here. No change in the insulating behavior is observed in this compound even for magnetic field application to the tune of 7 T. (b) A very weak field dependence in similar measurements on the metallic phase of Y$_{0.5}$Ca$_{0.5}$IrO$_{3}$ is observed. However, no quantitative analysis can be carried out as the resulting magnetoresistance is indistinguishable form the background. (c) \& (d) Similar measurements on stoichiometric  BaIrO$_{3}$ and Y$_{0.5}$Ba$_{0.5}$IrO$_{3}$ do not exhibit any field dependence of resistivity either. However, a sharp upturn in resistivity at $T$$\simeq$ 175 K is observed in both compounds.
} \vspace{-4mm}
\end{figure}

Electrical resistivity as a function of temperature for various chemical doping percentages of Y$_{x}$Ca$_{1-x}$IrO$_{3}$ are plotted in Fig. 1a. As we can see, CaIrO$_{3}$ is a strong insulator in stoichiometric composition. A small but observable change in slope of the resistivity curve around $T$$\simeq$~110 K is identified with an antiferromagnetic (AFM) transiton in this compound (as discussed below in latter paragraphs). After gradual substitution of Ca by Y, a flat plateau in electrical resistivity is found to develop below $\simeq$~120 K at x = 0.3, before ushering into completely metallic state at x = 0.5. This insulator-to-metal transition is described in a qualitative phase diagram of temperature vs. chemical doping percentage in Fig. 1b. The insulator-metal transition temperature for various yttrium doping percentages is estimated from the cusp in the resistivity data for that sample. In order to gain more information about electronic dimension, electrical resistivity data is fitted using variable range hopping (VRH) model.\cite{Cava} Fitting of electrical resistivity data below the magnetic transition temperature using VRH model, $\rho$($T$) = $\rho$$_{o}$~exp($T$$_{o}$/$T$$^{1/d+1}$), where $T$$_{o}$ is the magnetic transition temperature and $d$ is the dimensionality of hopping, reveals a fractional dimension of $d$ = 2.4 in CaIrO$_{3}$, Fig. 1c. Similar analysis of electrical data of BaIrO$_{3}$ demonstrates three dimensional electronic pattern, Fig. 1d. The quasi-2D electronic behavior in CaIrO$_{3}$ is different from previous assertion of quasi-1D magnetic behavior in this system.\cite{Bogdanov} Apparently, these two phenomena are not coupled to each other. Fractional dimensionality in a system also suggests a path-like electrical transport,\cite{Gordon} compared to surface-like propagation of charge carrier in integral dimension systems. Since path-like electrical transport derives from the underlying fractal-type chemical structure,  the disparity in electrical dimensions can be attributed to the arrangement of the IrO$_{6}$ octahedra surrounding alkaline earth ions in these compounds. 

\begin{figure}
\centering
\includegraphics[width=8.5 cm]{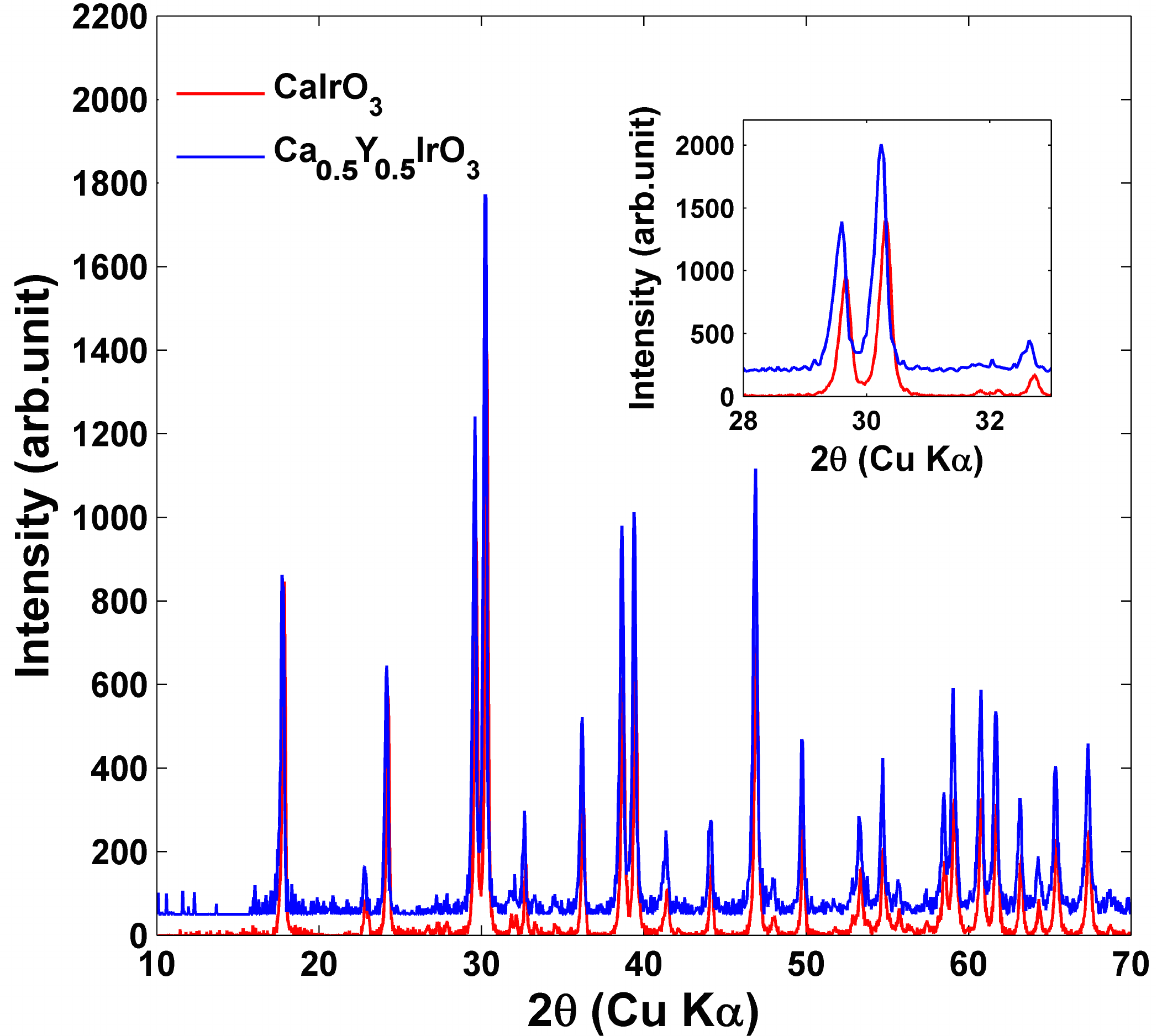} \vspace{-4mm}
\caption{(color online) Powder X-ray diffraction data of CaIrO$_{3}$ and Y$_{0.5}$Ca$_{0.5}$IrO$_{3}$. A very small shift in peak positions of Y$_{0.5}$Ca$_{0.5}$IrO$_{3}$ with respect to CaIrO$_{3}$, as highlighted in the inset, suggests that the underlying crystal structure is preserved. Similar behavior is observed in yttrium doping of BaIrO$_{3}$. 
} \vspace{-4mm}
\end{figure}

Yttrium substitution of Ba in BaIrO$_{3}$ does not change the insulating behavior found in the stoichiometric composition, as noted earlier. In Fig 2, characteristic plots of electrical resistivity as a function of temperature at various applied magnetic field values are depicted for two different compositions (x = 0 and x = 0.5) of CaIrO$_{3}$ and BaIrO$_{3}$. It is immediately noticed that the magnetic field application does not affect electrical resistivity in either of the compounds. Basically, no magnetoresistance, distinguishable form the background, is recorded. A downward cusp in the electrical data of BaIrO$_{3}$, Fig. 2c, at $T$ $\simeq$ 175 K is most likely associated to the charge density wave (coexisting with weak ferromagnetism), as recent reports have suggested.\cite{Cao1,Laguna,Maiti} The cusp is more pronounced in Y$_{0.5}$Ba$_{0.5}$IrO$_{3}$, which suggests stronger coupling between the charge density wave and the underlying crystal structure in this system. Yttrium susbstitution of alkaline earth ions does not alter the crystal structure drastically. For illustration purposes, X-ray diffraction patterns of CaIrO$_{3}$ and Y$_{0.5}$Ca$_{0.5}$IrO$_{3}$ are plotted in Fig. 3, where expected small shifts in the peak positions of Y$_{0.5}$Ca$_{0.5}$IrO$_{3}$ with respect to CaIrO$_{3}$ are observed. Lattice parameters for both sets of compounds are tabulated in Table 1. As we can see, the net change in lattice parameters is less than 2\%. Hence, the crystal structure remains intact even at 50\% substitution of Ba (Ca) by yttrium.

Next, we have performed detailed ac magnetic susceptibility measurements on both sets of compounds. AC susceptibility measurements provide information about static and dynamic properties of correlated spins at the same time. Net ac susceptibility $\chi$ ($T$) is written as: $\chi$($T$) = $\chi$$^{'}$ ($T$) + $i$ $\chi$$^{"}$ ($T$), where real $\chi$$^{'}$ represents static magnetic behavior and the imaginary $\chi$$^{"}$ provides information about dynamic magnetic properties or, damping of magnetic fluctuations in a system. Static and dynamic susceptibilities at two characteristic frequencies (10$^{3}$ and 10$^{4}$ Hz) for two doping percentages, x = 0 and 0.5, of Y$_{x}$Ca$_{1-x}$IrO$_{3}$ and Y$_{x}$Ba$_{1-x}$IrO$_{3}$ are plotted in Fig. 4. In CaIrO$_{3}$, a very sharp peak develops around $T$$_{N}$$\simeq$110 K in both static and dynamic susceptibilities. The location of the $\chi$$^{'}$ peak (in temperature) is consistent with a small change in slope of the resistivity profile, thus can be attributed to the onset of long range AFM order. At the same time, an equally sharp peak in dynamic susceptibility hints about resonance of paramagnetic ions with a single relaxation time that coincides with the onset of long range AFM order in this compound.\cite{Wu} Another noticeable feature in Fig. 4a and 4b is the lack of frequency dependence of peak location of static and dynamic susceptibilities, which rules out any spin freezing behavior or, the presence of multiple relaxation times among small paramagnetic clusters.\cite{Young} Magnetic measurements on Y$_{0.5}$Ca$_{0.5}$IrO$_{3}$ reveal an upturn (in addition to the sharp peak in $\chi$$^{'}$, perhaps weaker compared to CaIrO$_{3}$) in both static and dynamic susceptibilities at low temperature, Fig. 4c and 4d. While the gradual increment in susceptibilities at low temperatures can be related to a new dynamic magnetic order, diminishing peak intensity at higher frequencies in dynamic susceptibilities (Fig. 4d) hints of magnetic fluctuations at very low energy in this compound. Similar magnetic measurements of BaIrO$_{3}$ confirm the onset of weak ferromagnetism at $T$$\simeq$175 K, as evidenced by slightly broader peaks as a function of temperature in $\chi$$^{'}$ and $\chi$$^{"}$, Fig. 4e and 4f. A broad peak in static susceptibility is often associated to the formation of small magnetic clusters in a system. The peak value of static susceptibility in BaIrO$_{3}$ is at least an order of magnitude smaller than its calcium counterpart. Together these observations constitute an argument about the presence of small cluster-type ferromagnetic phenomenon in this compound. Yttrium substitution of Ba in Y$_{0.5}$Ba$_{0.5}$IrO$_{3}$ removes the broad feature in susceptibilities peaks of BaIrO$_{3}$, as shown in Fig. 4g and 4h. In addition to a sharp peak, an upturn in static susceptibility at low temperature is also observable in this case. Similar observations are recorded in the magnetic susceptibility measurements of other yttrium doping percentages of barium and calcium iridates.

\begin{table}
\caption{\label{OMEGA} Lattice parameters and angles, as derived from the refinement of powder X-ray diffraction data, for both sets of compounds.}
\begin{tabular}{| c | c | c | c | c| c| c|}
\hline 
& a ($\AA$) & b ($\AA$) & c ($\AA$) & $\alpha$(degree) & $\beta$ & $\gamma$ \\
\hline
CaIrO$_{3}$ & 3.145 & 9.863 & 7.297 & 90 & 90 & 90 \\
\hline
Y$_{0.5}$Ca$_{0.5}$IrO$_{3}$ & 3.148 & 9.875 & 7.30 & 90 & 90 & 90 \\
\hline
BaIrO$_{3}$ & 10.012 & 5.762 & 15.178 & 90 & 103.25 & 90 \\
\hline
Y$_{0.5}$Ba$_{0.5}$IrO$_{3}$ & 10.006 & 5.752 & 15.178 & 90 & 103.28 & 90 \\
\hline
\end{tabular}
\end{table}

\begin{figure}
\centering
\includegraphics[width=8.7 cm]{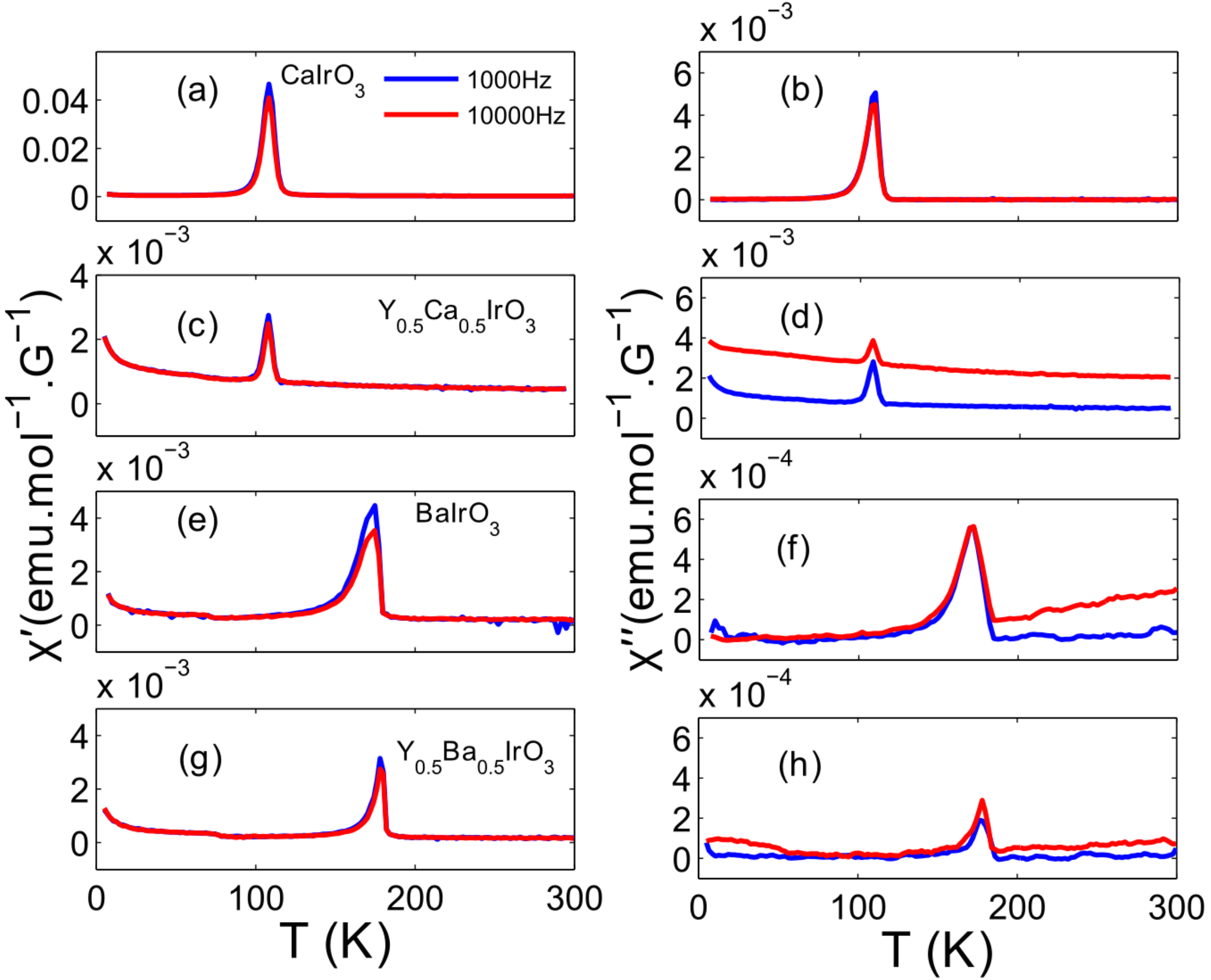} \vspace{-4mm}
\caption{(color online) AC susceptibility measurements of x = 0 and 0.5 chemical dopings of Y$_{x}$Ca$_{1-x}$IrO$_{3}$ and Y$_{x}$Ba$_{1-x}$IrO$_{3}$. (a) \& (b) Static ($\chi$$^{'}$) and dynamic ($\chi$$^{"}$) susceptibilites of CaIrO$_{3}$ as a function of temperature at two different characteristic frequencies (1000 and 10000 Hz). A sharp peak in $\chi$$^{'}$ around $T$$_{N}$$\simeq$ 110 K is attributed to the onset of long range AFM order in this compound. No frequency dependence of the peak position rules out any glassiness. A sharp frequency-independent peak is also observed in dynamic susceptibility, $\chi$$^{"}$, around the same temperature, indicating strong magnetic resonance associated to the AFM order in CaIrO$_{3}$. (c) \& (d) In addition to a sharp peak around $T$$\simeq$ 110 K, a gradual increment in both static and dynamic susceptibilites of Y$_{0.5}$Ca$_{0.5}$IrO$_{3}$ is observed below $T$$\simeq$50 K. Also observable is the weakness in the dynamic susceptibility at higher frequency. ($\chi$$^{"}$ at different frequencies are separated by 0.002 emu.mol$^{-1}$.G$^{-1}$ to highlight this observation.) (e) \& (f) A broad peak as a function of temperature, centered around $T$$\simeq$ 175 K, is observed in the ac susceptibility measurements of BaIrO$_{3}$. (g) \& (h) Unlike BaIrO$_{3}$, the peak in $\chi$$^{'}$ is sharper in Y$_{0.5}$Ba$_{0.5}$IrO$_{3}$. Also an upturn in $\chi$$^{'}$ is observed below $T$$\simeq$100 K. Once again, there is little or no frequency dependence of susceptibility peak intensity or location as a function of temperature.
} \vspace{-4mm}
\end{figure}

An important difference between the stoichiometric composition and yttrium substituted compounds lies in the observation of an upturn in static susceptibility at low temperature, which in principle suggests the occurence of new magnetic order in Y$_{x}$M$_{1-x}$IrO$_{3}$ (M = Ca, Ba). While it is not possible to rule out the development of another minor magnetic phase due to yttrium doping, powder X-ray data confirms high purity of Y$_{x}$M$_{1-x}$IrO$_{3}$ for all chemical doping percentages (Fig. 3). We also note that the electrical resistivity exhibits little or no field dependence in any of these materials. In Y$_{x}$Ba$_{1-x}$IrO$_{3}$, however, two notable behaviors are observed: first, a sharp downward cusp at the characteristic temperature, $T$$\simeq$175 K, and second, steep rise in resistivity below $T$$\simeq$80 K. While former phenomenon is found to be associated to the development of coexisting charge density wave and weak ferromagnetism,\cite{Cao2,Maiti,Nakano,Lindsay} latter observation is reported for the first time in this letter and requires further investigations. 

Our studies of Y$_{x}$$M$$_{1-x}$IrO$_{3}$, where $M$ = Ca, Ba, using electrical and magnetic measurements show unambiguous new results on the presence (absence) of insulator-to-metal transition in post-perovskite (perovskite) iridium oxide compounds. In stoichiometric compositions, both CaIrO$_{3}$ and BaIrO$_{3}$ are Mott insulators with ($t$$_{2g}$)$^{5}$ electronic configuration, leading to $j$$_{eff}$ = 1/2 state. The unexpected Mott insulating ground state, in fact, arises due to an interaction between the spin-orbit coupling ($\zeta$$\simeq$0.5 eV) and the cubic crystal field (10Dq $\simeq$ 3 eV).\cite{Zhang,Sala1,Boseggia} In absence of the strong spin-orbit coupling, these compounds should exhibit metallic properties. In addition to these two terms, a tetragonal distortion, $\Delta$ ($\simeq$ -0.01 eV), to the cubic crystal field is suggested to play an important role.\cite{Boseggia,Sala1} The Mott insulating $j$ = 1/2 state is possible as long as $\Delta$ $\ll$$\zeta$. This additional term is especially important in CaIrO$_{3}$, which undergoes significant structural distortion.\cite{Hirai} The tetragonal distortion $\Delta$ ($\simeq$ -0.7 eV) in CaIrO$_{3}$ is found to be stronger than the spin-orbit coupling strength $\zeta$ ($\simeq$ 0.5 eV), as recent measurements of resonant inelastic X-ray scattering (RIXS) have demonstrated. It is still debated whether CaIrO$_{3}$ exhibits a $j$ = 1/2 ground state or not. Nonetheless, the structural distortion giving rise to stronger $\Delta$ perhaps diminishes the spin-orbital interaction strength in CaIrO$_{3}$, thus an easier candidate for manipulation of electronic properties. It is speculated that the structural distortion also leads to the fractional electronic dimension, indicating path-like electrical transport, in this compound. Unlike CaIrO$_{3}$, BaIrO$_{3}$ does not undergo strong structural distortion. While not much is known about various competing energy terms in this compound, it is reasonable to argue that the robustness against the crystal structure distortion makes it unlikely to exhibit insulator-to-metal transition due to additional charge carrier. The yttrium substitution of alkaline earth metals does not seem to affect the magnetic transition temperature in Y$_{x}$$M$$_{1-x}$IrO$_{3}$. It suggests that the electrical transport and magnetic properties are two parallel phenomena, where the structural distortion influences former strongly. More experimental and theoretical research are highly desirable to further understand these anomalous properties. We also note that if doping of Mott insulator is the physics of high temperature superconductivity, as argued by some researchers in the case of cuprate superconductors,\cite{Lee} then our results provide a new platform to test this hypothesis.

We want to thank W. Montfrooij for helpful discussion. DKS acknowledges support from the University of Misosuri Research Board and IGERT research program, funded by NSF under grant number DGE-1069091.

\clearpage

\end{document}